\documentclass[preprint,5p,twocolumn]{elsarticle}

\usepackage{mathrsfs,amsmath,amsthm,amssymb,braket,color,verbatim,adjustbox,multirow,enumerate,textcomp,gensymb,subfigure,graphicx,diagbox,pifont,tabulary,booktabs,stackengine,epstopdf,dcolumn,makecell,listings,bbm,mathtools,autobreak,slashed,float,array,colordvi,xfrac}
\usepackage{lineno}
\modulolinenumbers[5]
\usepackage{hyperref}
\hypersetup{colorlinks,bookmarksopen,bookmarksnumbered,citecolor=[rgb]{0,0.35,0.5},linkcolor=[rgb]{0,0.35,0.5},urlcolor=[rgb]{0,0.35,0.5},pdfstartview=FitH,linktocpage}

\definecolor{eprintLinks}{rgb}{0,0.35,0.5}
\definecolor{journalLinks}{rgb}{0,0.35,0.5}
\newcommand{\MYhref}[3][blueLinks]{\href{#2}{\color{#1}{#3}}}

\def\g{\gamma}
\def\dla{\langle\!\langle}
\def\dra{\rangle\!\rangle}
\def\Ms{M_{s}}
\def\Md{M_{D}}
\def\gs{g_{s}}
\def\O{{\cal O}}
\def\pd{\partial}
\def\d{\delta}
\def\D{\Delta}
\def\ls{\ell_{s}}
\def\alp{\alpha^{\prime}}
\def\la{\langle}
\def\ra{\rangle}
\def\Elv{E_{\textrm{LV}}}
\def\Ep{E_{\textrm{Pl}}}
\def\sd{\textrm{d}}

\journal{PLB, published as Phys. Lett. B 869 (2025) 139823, \href{https://doi.org/10.1016/j.physletb.2025.139823}{doi:10.1016/j.physletb.2025.139823}}

\begin{document}
\hypersetup{bookmarksopen,bookmarksnumbered,citecolor=[rgb]{0,0.35,0.5},linkcolor=[rgb]{0,0.35,0.5},urlcolor=[rgb]{0,0.35,0.5},linktocpage}
\biboptions{numbers,sort&compress}

\begin{frontmatter}

\title{Shower formation in the presence of a string-inspired foam in space-time}

\author[pku]{Chengyi Li\corref{em}}
\cortext[em]{Corresponding author \ead{lichengyi@pku.edu.cn}}

\address[pku]{School of Physics, Peking University, Beijing 100871, China}

\begin{abstract}
It was recently proposed that predictions of Lorentz-breaking space-time foam models from string theory may be compatible with the suggestion of light-speed variation from gamma-ray burst studies. Our analysis of foam-modified kinematics shows that despite the subluminal photon velocities explaining photon time delays one may, and in certain circumstances does, keep intact the electromagnetic showers essential for the detection and identification of cosmic photons. In contrast to other~(mostly phenomenological) approaches to Lorentz violations with modified dispersions leading to drastic changes on the formation length for the cascade development in the atmosphere and in detectors, there is the possibility that the dispersion effect in the present string foam model avoids such modifications and the theory naturally escapes the shower formation constraints from recent observations.
\end{abstract}


\end{frontmatter}

\section{Introduction}

We know that Lorentz symmetry is crucial in fundamental physics and any idea and/or experimental indication of its violation which might characterize some quantum gravity~(QG) theories~\cite{Li:2025stf}, should be treated with care. Models entailing Lorentz violation~(LV) generically predict energy-dependent speed variations for relativistic probes provided the space-time probed at a small enough scale could appear complicated---something akin in complexity to a turbulent froth that Wheeler dubbed \emph{space-time foam}~\cite{Wheeler:1998stf}. Energetic $\g$-rays from cosmic sources like gamma-ray bursts~(GRBs) provide a window to search for such effects through the so-induced time delays~\cite{Amelino-Camelia:1997ieq,Jacob:2008bw}. Specifically, the authors of~\cite{Xu:2016zxi,Xu:2016zsa} analyze a dataset consisting of 14 multi-GeV photons from 8 Fermi GRBs and upon incorporating a constant intrinsic lag at the source~(see also~\cite{Shao:2009bv,Zhang:2014wpb,Amelino-Camelia:2016ohi,Xu:2018ien,Liu:2018qrg,Chen:2019avc,Zhu:2021pml,Zhu:2022usw}) in the observed time lags, they suggest a light-speed variation, $v(E)=c(1-E/\Elv)$, with the LV scale, $\Elv\sim 3\times 10^{17}$~GeV. This corresponds to a subluminal dispersion relation,
\begin{equation}
\label{eq:1}
E^{2}\simeq c^{2}k^{2}\Bigl(1-s\frac{ck}{\Elv}\Bigr),
\end{equation}
with the sign factor $s=+1$ and $\Elv$ approaching the order of Planck-scale $\Ep\sim 10^{19}$~GeV. We shall use the units so that $c=\hslash=1$ with $c$ the constant speed of light. Recently, the Monte Carlo~(MC) analysis~\cite{Song:2025akr} implies the necessity of introducing energy-dependent intrinsic delays~\cite{Song:2024and} in order for 3 more remarkable events across GeV to TeV bands to be robustly fitted within the same physics scenario~(\ref{eq:1})~\cite{Song:2025qej}. With the full Bayesian parameter estimation strategy employed, a recent work in~\cite{Song:2025myx} finds that the null hypothesis of dispersion-free vacuum $E=k$, or, equivalently, the constant light speed $v=1$ is rejected at a significance level of $3.1\sigma$ or higher.

While one might consider earlier studies~\cite{Shao:2009bv,Zhang:2014wpb,Amelino-Camelia:2016ohi,Xu:2016zxi,Xu:2016zsa,Xu:2018ien,Liu:2018qrg,Chen:2019avc,Zhu:2021pml,Zhu:2022usw} not so reliable~(due to imperfect knowledge of the radiation mechanism of GRBs) and LV hypothesis too premature the statistical analysis involving MC simulation has indicated the robustness of the finding. Besides conventional astrophysical explanations like matter effects which can mimic LV~\cite{Zhang:2014wpb}, some exotic interpretations of the result were recently suggested~\cite{Li:2021gah,Li:2021tcw,Li:2022sgs}. These are based on string models~\cite{Ellis:1999uh,Ellis:2000sf,Ellis:2004ay,Mavromatos:2005bu,Ellis:2008gg,Mavromatos:2012ii,Li:2022sgs,Li:2023wlo,Li:2024crc,Li:2009tt} of space-time foam-induced refractive index \emph{in vacuo} $\eta$ which is subluminal, growing linearly with photon energy $\omega(=E)$ and, suppressed by a single power of the string mass scale, $\Ms$, viz.,
\begin{equation}
\label{eq:2}
\eta-1\simeq\omega/f(\Ms),\quad f(\Ms)\propto\Ms.
\end{equation}
Here the functional form of QG scale $f(\Ms)\eqqcolon M$ depends on foam types and $M\sim\Elv$, indicated by the above phenomenological finding. The important point to note here is that such explanations are also consistent with other stringent tests of LV, in particular constraints from $\g$-decay and birefringence effect~(for details see~\cite{Li:2021gah}).

The common wisdom regarding the impact of LV is that, although it affects light propagation in free space~(and possibly the thresholds of photoproduction collisions with soft photon of cosmic background lights) the effect of the modified dispersion relation on interactions relevant for the detection of high-energy emissions is negligible. Nonetheless, in~\cite{Kifune:1999ex}, the author first points this may not hold true: electromagnetic~(EM) interactions like pair creation in the detectors onboard satellites or Cherenkov process in ground-based observatories may be modified by LV. A later analysis further notes~\cite{Vankov:2002gt} that dispersion effect, of the type~(\ref{eq:1}), will imply unacceptable changes in the development of EM shower cascade. In particular, a strong suppression of photon ``decay'' into an $e^{\mp}$ pair in the Coulomb field of nuclei, the process mainly responsible for the formation of photon-initiated atmospheric showers~\cite{Risse:2007sd}, could be expected~\cite{Rubtsov:2012kb}. For PeV $\g$-rays at the Large High Altitude Air Shower Observatory~(LHAASO)~\cite{LHAASO:2021gok}, showers that are observed with no deviation from standard cascade theory have been used recently in~\cite{Satunin:2023yvj} to severely constrain the field theory of~\cite{Myers:2003fd}: $\Elv\gtrsim\O(10^{20})$~GeV for subluminal LV photons~(\ref{eq:1}). However, all of these conclusions are based on assuming energy conservation which may be invalid in the above-mentioned string-inspired ``foamy'' situations.

Thus, in this Letter we aim to discuss the influence from stringy space-time foam and the associated energy fluctuations on particle shower formation, which is still unknown, and further examine whether the obtained results from the phenomenological or, at most, the effective field theory approaches can still apply, once such fluctuations induced by the foam are present. Here, it is worth noticing that different formulations in this framework have suggested different modifications of the standard physics. For example, for the foam type in~\cite{Ellis:2008gg}, which does not entail deformed relations between energy and momentum, there is no mechanism for energy nonconservation despite the presence of subluminal photons. In this Letter, however for our purposes, we only focus on discussing QG effect originating from an adaption of the model recently formulated in our Ref.~\cite{Li:2022sgs} on shower development. It is shown that under certain conditions the formation length for pair creation and bremsstrahlung and thus qualitatively, their cross-section, are unaffected; then the light-speed variation would \emph{not} lead to drastic changes to photon showers in the context of such models and evades the above constraints.

\section{String foam and the shower formation}

In string models for space-time foam, the in-vacuo propagation of photons is modified as initially inspired from Liouville QG~\cite{Ellis:1992eh}, with their travel times delayed by amounts proportional to $\eta-1$~(c.f.~(\ref{eq:2})), due to D-brane space-time defects. These defects break Poincar\'e invariance, and their recoil during interactions with an open string representing a photon breaks local Lorentz invariance. They give space-time a foamy nature so the situations were termed \emph{D-foam}. The cosmological setup of D-foam is based on braneworlds moving in a higher-dimensional bulk space that contains a ``gas'' of D0-branes~(D-particles)~\cite{Ellis:2004ay} of type I/IIA strings. We consider it to be present in models of phenomenological interest~\cite{Ellis:1999uh,Ellis:2000sf,Ellis:2004ay,Mavromatos:2005bu,Ellis:2008gg,Mavromatos:2012ii,Li:2021gah,Li:2021tcw,Li:2022sgs,Li:2023wlo,Li:2024crc} as even if the theory admits no D-particles one has effective ones~\cite{Li:2009tt}. As our brane roams in the bulk, they cross it randomly, and collide with the photon resided on it and it is not possible in general to capture the process with local effective Lagrangians.

We have previously investigated stochastic effects of the foam in a model of this kind in~\cite{Li:2022sgs} which will violate CPT, and neutrinos may experience a so-induced CPT violation during propagation in this QG ``medium'', dictated by the dispersion relations:
\begin{equation}
\label{eq:3}
E_{\nu,\overline{\nu}}=E_{M}\mp\frac{1}{f(\Ms)}k^{2},
\end{equation}
where $E_{M}\equiv\sqrt{k^{2}+m_{\nu}^{2}}$ and $f(\Ms)=\frac{2\Md}{\zeta^{2}}$~($\zeta$ a parameter characterizing stochastic foam-modified neutrino propagation, $\Md\coloneqq\frac{\Ms}{\gs}$ is D-particle mass, $\gs\ll 1$ is~(weak) string coupling). The argument implying broken CPT for $\nu$'s and $\overline{\nu}$'s, hence to superluminal $\overline{\nu}$'s, applies only to fermions, as opposed to bosons like photons.

In order to discuss cosmic photon-induced showers let us first derive the photon dispersions, via considering the nontrivial momentum transfer $\D k_{i}\equiv rk_{i}$, in each collision of a photon off a D-(particle)-defect, which recoils at a velocity, $u$ with its components along the brane, $u_{i}=r\gs k_{i}/\Ms$~\cite{Ellis:1999uh,Ellis:2000sf,Ellis:2004ay}. We assume isotropic random recoil, as in~\cite{Li:2022sgs}, i.e., the momentum fraction variable $r$ vanishes on average $\dla r\dra=0$ and $\dla r^{2}\dra\eqqcolon\sigma^{2}<1$, where $\dla\cdot\cdot\cdot\dra$'s are statistical averages over a collection of quantum-fluctuating defects met by the photon. Here, the variance $\sigma$ plays a role similar in nature to the parameter $\zeta$ for neutrino species~($\sigma\neq\zeta$ in general).\footnote{The topologically nontrivial effects of the foam on particle propagation do not obey the principle of equivalence~\cite{Ellis:2003ua,Ellis:2003sd}, in the sense of being universal for all types of excitations of the Standard Model; moreover for reasons of charge conservation such a medium of defects is ``transparent'' for charged particles, such as electrons~(as indicated by experiment~\cite{Li:2022ugz}).} The metric surrounding the struck defect is of Finsler type $G_{\mu\nu}=\eta_{\mu\nu}+h_{\mu\nu}$ and $h_{0i}=u_{i}$ modifying the energy of the photon via $k^{\mu}k^{\nu}G_{\mu\nu}=0$. One obtains after averaging over foam particles the modified energy, $\omega\simeq k[1+\frac{1}{2}\sigma^{2}k^{2}/M_{D}^{2}]$, used in the kinematics: $\omega=\omega^{\prime}+\dla\Md u^{2}/2\dra$, where $\frac{\Ms}{2\gs}u^{2}$ is the kinetic energy of heavy defects in the case of nonrelativistic recoil, $u_{i}\ll 1$. The final dispersion relation found from $\omega^{\prime}$ in the leading approximation yields,
\begin{equation}
\label{eq:4}
\omega(k)\equiv\omega^{\prime}=k-\sigma^{2}\gs\frac{k^{2}}{2\Ms},
\end{equation}
here $\sigma^{2}$ a free parameter, depends on foam density $n_{D}$~(and microscopic details that cannot be determined quantitively presently). It is clear that this subluminal dispersion effect leads, through the appropriate group velocity~($v\coloneqq\pd\omega/\pd k$ assumed to hold), to a QG refractive index~(\ref{eq:2}) for photons: $v\simeq 1-\omega\sigma^{2}/\Md$, of the kind favored~\cite{Li:2021gah} by current time-lag studies of GRB photons~\cite{Shao:2009bv,Zhang:2014wpb,Amelino-Camelia:2016ohi,Xu:2016zxi,Xu:2016zsa,Xu:2018ien,Liu:2018qrg,Chen:2019avc,Zhu:2021pml,Zhu:2022usw,Song:2024and,Song:2025myx,Song:2025qej,Song:2025akr}. For a uniform D-defect density, at least for late eras of the Universe history where $\sigma$ is constant, an order of magnitude $\sigma^{2}\sim\O(1)$ is sufficient to reproduce the finding of light-speed variation mentioned earlier, assuming a Planckian D-particle mass $\sim 10^{18}$~GeV. In cases of lower $\Md$ the required $\sigma^{2}\ll 1$ is easily satisfied with more dilute foam populations.

It has been argued~\cite{Satunin:2023yvj} however that, due to the presence of the modified photon dispersion~(\ref{eq:1}) which, as we see, may originate from a string-inspired scenario~(c.f.~(\ref{eq:4})) the cross-sections of standard~(perturbative) processes are modified. In particular, any allowed interaction channel is suppressed for subluminal photons, since such LV photons lack energy compared to the momentum. Earlier as argued in~\cite{Vankov:2002gt} they may even suppress the cascade generated by cosmic $\g$-rays, causing the showers to be formed deeper in the atmosphere and in detectors. The absence of the effect in the measured spectrum sets restrictive~(shower formation) constraints on $\Elv$~\cite{Satunin:2023yvj}. However, these considerations also concern strict energy--momentum conservation~(assumed in prior works), not only modified dispersion relation.

In general, in the space-time foam picture energy is conserved only in a statistical sense~(see the review~\cite{Sarkar:2002mg}), however, even this can be violated due to unobserved degrees of freedom~(from the point of view of a low-energy braneworld observer) associated with the D-particle recoil in the model of~\cite{Li:2022sgs}. Indeed, there are quantum fluctuations $\d E_{D}$ in the total energy in particle reactions~\cite{Ellis:2000sf}. For a three-body decay process in the recoil foam adapted to the situation here discussed, the energy of the incoming particle $E$ fluctuates, due to the presence of stochastic D-foam, by the amount of $(\varsigma_{I}/\Md)p^{2}$ where $p$, the momentum of the particle emitted during the decay of a particle of effective energy $E+\d E_{D}$. The proportional factor $\varsigma_{I}$ expressing the intensity of such fluctuations can vary independently of the foam dispersion factor $\sigma$ and values $\varsigma_{I}>0$, as in principle permitted by the theory, will be seriously considered here. Now we are going to see how this affects the conclusion drawn from other LV approaches~\cite{Vankov:2002gt,Myers:2003fd}, yielding unacceptable anomalies in the formation of cosmic $\g$-ray showers. As we shall see, the peculiarity of this foam in resulting in such energy violations is crucial for accommodating the result of light-speed variation, without conflicting with the constraints from shower measurements.

At the Earth photons initiate almost purely EM showers via pair creation and bremsstrahlung~\cite{Risse:2007sd}. Additional processes are important only at highest energy~($10^{19}-10^{20}$~eV) so for GeV to sub-PeV $\g$-rays we ignore them for simplicity except where noted. Lacking at present a complete theory of the matter/D-foam interactions we do not deal with the cross-sections directly and shall restrict ourselves to calculate the length $\lambda_{f}$, over which the processes are considered taking place according to uncertainty principle, that is formation~(or coherent) length~\cite{Klein:1998du}.

First, let us consider pair creation or Bethe--Heitler~(BH) process, which dictates the first reaction of astrophysical $\g$-rays in the atmosphere $\g Z\rightarrow Zee^{+}$. This can be viewed as photon colliding on a virtual photon $\g^{\ast}$, from the Coulomb field of the nucleus of charge $Z$. The momentum exchanged with $\g^{\ast}$ is,
\begin{equation}
\label{eq:5}
\D q_{z}=k-p-p^{\prime},
\end{equation}
where $p$, $p^{\prime}$ indicate the outgoing momenta of electron and positron and photon's motion is in $z$ axis, with its momentum $k$. The nucleus' Coulomb field has a larger component perpendicular to $z$ axis, leading to a transverse momentum kick $\D q_{\perp}\sim\O(m_{e})$~($m_{e}$, the electron rest mass), while the longitudinal momentum $\D q_{z}$ is small, $\D q_{z}\ll k,\D q_{\perp}$~\cite{Rubtsov:2012kb}, due to the photon's high energy $\omega\gg 2m_{e}$. For this regime one can ignore collision angles. In the presence of string D-foam, one should make use of the dispersion including the foam induced LV~(\ref{eq:4}). For electron~(positron) its law of dispersion is preserved as noted, for gauge
invariance~\cite{Ellis:2003ua,Ellis:2003sd}, so we have $E^{2}-m_{e}^{2}=p^{2}$ and then,
\begin{align}
\label{eq:6}
&p=\sqrt{E^{2}-m_{e}^{2}}\simeq E\Bigl(1-\frac{m_{e}^{2}}{2E^{2}}\Bigr),\\
\label{eq:7}
&p^{\prime}\simeq E^{\prime}\Bigl(1-\frac{m_{e}^{2}}{2E^{\prime 2}}\Bigr),
\end{align}
where $E^{\prime}$ is the positron energy. Now, as the energy fluctuation of the particle system due to the foam is considered, $E^{\prime}=\omega-E+\d E_{D}$. It is worth stressing that in D-particle models, LV acts as a back-reaction effect of recoil on space-time, thus it can affect only real particles. Virtual photons are not able to ``see'' special QG configuration~(vice versa) so this process is simplified to a three-body ``decay'' in the present model. So we have,
\begin{equation}
\label{eq:8}
\D q_{z}=\frac{m_{e}^{2}}{2E}+\frac{m_{e}^{2}}{2(\omega-E+\d E_{D})}+\sigma^{2}\frac{k^{2}}{2\Md}-\d E_{D}.
\end{equation}
Following from above discussion of the energy violation, it is given by
\begin{equation}
\label{eq:9}
\d E_{D}\simeq\frac{\varsigma_{I}}{\Md}p^{2}\simeq\varsigma_{I}\frac{k^{2}}{4\Md},
\end{equation}
where $p\simeq\frac{k}{2}$, i.e., momentum still balances approximately between the $ee^{+}$ pair. Substituting~(\ref{eq:9}) to Eq.~(\ref{eq:8}) the latter becomes,
\begin{align}
\label{eq:10}
\D q_{z}\simeq\frac{m_{e}^{2}}{2E}&-\varsigma_{I}\frac{k^{2}}{4\Md}+\sigma^{2}\frac{k^{2}}{2\Md}\nonumber\\
&+\frac{m_{e}^{2}}{2(k-E)}\Bigl[1-\frac{\gs(\varsigma_{I}-2\sigma^{2})}{4\Ms}\frac{k^{2}}{k-E}\Bigr].
\end{align}
For string scale in traditional string theories assumed here, $\Ms\sim 10^{18}$~GeV~(order of the reduced Planck mass) $m_{e}\ll k,E\ll\Md$ at high energy, $\O(m_{e}^{2}/\Ms)$ term in Eq.~(\ref{eq:10}) is negligible. Then we find,
\begin{equation}
\label{eq:11}
\D q_{z}\simeq\frac{m_{e}^{2}\omega}{2E(\omega-E)}+(2\sigma^{2}-\varsigma_{I})\gs\frac{\omega^{2}}{4\Ms},
\end{equation}
where, to leading order, $k\simeq\omega$. If $\varsigma_{I}$ vanishes, this formula coincides with the one derived in models that simply entail subluminal LV~($\sigma^{2}>0$). It would recover the conventional result as $n_{D}\rightarrow 0$.

We come to discuss the formation length, $\lambda_{f}$, dictated by Heisenberg's uncertainty principle. While the latter suffers a modification, due to the existence of the minimal~(string) length $\ls=1/\Ms$ in target space-time~\cite{Veneziano:1986zf},
\begin{equation}
\label{eq:12}
\D x\sim\frac{1}{\D p}+\alp\D p,
\end{equation}
(where $\ls^{2}=\alp$ is the string's Regge slope) to leading order in $\Md^{-1}$, it suffices to use the usual relation which requires,
\begin{equation}
\label{eq:13}
\lambda_{f}=\frac{1}{\D p_{z}},
\end{equation}
which is just the wavelength of the exchanged virtual photon and the distance over which the amplitudes contribute coherently to the cross-section, that scales linearly with $\lambda_{f}$ as a result. To quantify possible changes of the process due to different factors, it is convenient to define a suppression factor giving the differential cross-section modified by foam effects relative to Bethe--Heitler,
\begin{align}
\label{eq:14}
S&=\frac{\sd\tilde{\sigma}/\sd\D q_{\perp}}{\sd\sigma_{\textrm{BH}}/\sd\D q_{\perp}}=\frac{\lambda_{f}}{\lambda_{f0}}\nonumber\\
&\simeq\frac{1}{1+\omega^{3}\gs(2\sigma^{2}-\varsigma_{I})/(8\Ms m_{e}^{2})}
\end{align}
where, as in the Lorentz-invariant case, $\lambda_{f0}=\frac{2E(\omega-E)}{m_{e}^{2}\omega}$. In Eq.~(\ref{eq:14}), we have taken the electron and positron energies, $E\simeq\omega-E\simeq\frac{1}{2}\omega$, for reasons mentioned above, then, both the modified and foam-free formation lengths are maximal due to minimal $\D q_{z}$'s:
\begin{equation}
\label{eq:15}
\D q_{z}^{\min}=\frac{1}{\lambda_{f0}^{\max}}-\frac{\omega^{2}(\varsigma_{I}-2\sigma^{2})}{4\Md},
\end{equation}
where $\lambda_{f0}^{\max}=\omega/(2m_{e}^{2})$. The ratio~(\ref{eq:14}) depends on hierarchies among both parameters for the foam, not only values of $\sigma^{2}$ describing light-speed variation. If $\sigma^{2}>\frac{1}{2}\varsigma_{I}$ the foam causes suppression $S<1$~(c.f. $\lambda_{f0}>\lambda_{f}$), as in generic subluminal LV models, increasing the average depth of shower maximum, $\la X_{\max}\ra$, compared to the standard expectation, $\la X_{\max}\ra\simeq X_{\textrm{atm}}$~(the atmospheric depth), because, $X_{\max}$ is proportional to the product of logarithm of primary energy $\ln\omega$ and the radiation length. Notably, suppression rapidly increases, and the cross-section $\tilde{\sigma}$ drops, with increasing $\omega$; the mean interaction depth $\la X_{0}\ra$, increases proportionally, $\la\tilde{X}_{0}\ra=m/\tilde{\sigma}$~(where $m$ is the average mass of the atoms of the air, typically, nitrogen), as follows,
\begin{equation}
\label{eq:16}
{\la\tilde{X}_{0}\ra}=\frac{\sigma_{\textrm{BH}}}{\tilde{\sigma}}\la X_{0}\ra\propto S^{-1}\la X_{0}\ra,
\end{equation}
where $\la X_{0}\ra\simeq 50~\textrm{g}\cdot\textrm{cm}^{-2}$ with $\sigma_{\textrm{BH}}\approx 0.51$~b in air.
Then, the probability for a $\g$-ray to develop a shower and then be detected as penetrating through the Earth's atmosphere is given in~\cite{Satunin:2023yvj} and when adapted to our case reads
\begin{align}
\label{eq:17}
P&=\int_{0}^{X_{\textrm{atm}}}\sd X_{0}\frac{e^{-X_{0}/\la\tilde{X}_{0}\ra}}{\la\tilde{X}_{0}\ra}=1-e^{-X_{\textrm{atm}}/\la\tilde{X}_{0}\ra}\nonumber\\
&\simeq 1-\exp\biggl[-\Bigl(\frac{X_{\textrm{atm}}}{50~\textrm{g}\cdot\textrm{cm}^{-2}}\Bigr)\frac{8m_{e}^{2}\Ms}{\gs\omega^{3}(2\sigma^{2}-\varsigma_{I})}C\biggr],
\end{align}
here $C$ denotes some unknown and model-dependent factor and the strong suppression limit $\Md\ll \omega^{3}/[8m_{e}^{2}(2\sigma^{2}-\varsigma_{I})]$ is assumed. This leads to suppressed photon flux that can be tested against registered $F_{\textrm{reg}}$ by detectors from a given source:
\begin{equation}
\label{eq:18}
\frac{\sd\tilde{F}_{\textrm{reg}}}{\sd\omega}=P\cdot\frac{\sd F_{\textrm{reg}}}{\sd\omega}.
\end{equation}
It is from these considerations that interesting constraints on subluminal-photon linear LV were derived recently from the LHAASO data for TeV--PeV photon-induced air showers~\cite{LHAASO:2021gok}. Unlike the linear LV model of~\cite{Vankov:2002gt,Myers:2003fd}, however, in our case, such limits are not imposed on dispersion parameter $\sigma^{2}$ which is related to $\Elv$ in Eq.~(\ref{eq:1}) but for $2\sigma^{2}-\varsigma_{I}$. Naively, one could use the result they derive to make an order of magnitude estimate in the parameters of the present model:\footnote{We notice that for the result of~\cite{Satunin:2023yvj}, the effect of LV-modified pair creation on soft cosmic background photons leading to the increase of the flux is also taken into account, whilst it is subleading compared to the shower formation effects~(also for priorly studied spectra~\cite{Satunin:2019gsl,Rubtsov:2016bea}). However, the existence of such modifications in the present model has not been clarified, hence temporally ignored here and is left for future investigation.}
\begin{equation}
\label{eq:19}
2\sigma^{2}-\varsigma_{I}\sim 10^{-2}.
\end{equation}
By itself this imposes no constraint, since~(for traditionally high string scales assumed earlier) the most probable order of $\varsigma_{I}$ will also be $\O(1)$, albeit it implies $\varsigma_{I}\approx 2\sigma^{2}$. In fact, as we have mentioned, in such D-particle/string foam models, both parameters of the foam are adjustable, even the string scale $\Ms$, which is, at least at present, considered arbitrary. Intriguingly~(\ref{eq:19}) is saturated on condition that $\sigma=\sqrt{\varsigma_{I}/2}$ in which case no suppression is expected at all for photon's first interaction in the medium $S=1$ without anomaly for registration of cosmic photons. It is the energy fluctuations caused by the quantum D-foam in this approach that offset the effect of the dispersion relation.

The other effect that opposes those from general and/or field-theoretic description of subluminal LV photons is that if $\varsigma_{I}>2\sigma^{2}$, Eq.~(\ref{eq:14}) turns into enhancement which implies that, the formation length increases instead of being shortened. And the factor $S$ even becomes infinite at the critical energy $\omega_{\textrm{cr}}=\bigl[\frac{8\Md m_{e}^{2}}{(\varsigma_{I}-2\sigma^{2})}\bigr]^{1/3}$ and negative above it, while the situation could be compensated by suppression factors associated with interactions that can disrupt coherence over such a longer length. The effects, that increase $\D q_{z}$ and reduce $\lambda_{f}$, include, for example, the Landau--Pomeranchuk--Migdal effect~(see, e.g.,~\cite{Klein:1998du,Risse:2007sd}) due to multiple scattering centers. The so-reduced BH cross-section delays the development of the EM cascade initiated by a photon competing with foam effects which decrease now $\la X_{0}\ra$ and $\la X_{\max}\ra$. It is worth noting that the length of shower development $\D X$ shifting $X_{\max}$ as $X_{\max}=X_{0}+\D X$ is assumed unmodified in the above discussion, since the secondary interactions in the cascade are less energetic than the primary one. While, in principle, it may be affected by ``exotic'' physics studied here, not accounting for this leaves the estimate~(\ref{eq:19}) more conservative.

Nonetheless, let us finally give some heuristic arguments why bremsstrahlung may not be similarly affected. We use again the previous quantum mechanical arguments to find the formation length. In this case, the momentum transfer in the longitudinal~($+z$) direction is,
\begin{equation}
\label{eq:20}
\D q_{z}=p-p^{\prime}-k,
\end{equation}
where $p$ and $p^{\prime}$ stand now for the electron momenta before and after the interaction, and the photon emission angle is neglected for highly relativistic electron $E\gg m_{e}$. One difference herein is that, unlike pair creation where the energy partition in the produced $ee^{+}$ pair is most likely symmetric as we assumed above,\footnote{It was shown~\cite{Rubtsov:2012kb} that LV may enhance the leading particle effect as the cross-section is peaked at $E\sim\omega$ and $\omega-E\sim\omega$. But, there is no good reason to expect the same for D-foam cases where, unlike the model studied therein, to first order in $1/\Ms$, the matrix element for any process agrees with the one in the special-relativistic field theory so we consider symmetric pairs as an illustration.} the probability of asymmetric share is significant in bremsstrahlung. The resulting spectrum is dominated by photons emitted at very low energy~($\omega\ll E$) and diverges as $\omega\rightarrow 0$~($1/\omega$ dependence). Then the effects of foam for such configurations of soft bremsstrahlung photons are negligible, compared to the first two conventional, leading terms as in~(\ref{eq:8}):
\begin{equation}
\label{eq:21}
\D q_{z}\sim\frac{m_{e}^{2}\omega}{2E(E-\omega)},
\end{equation}
where Lorentz-invariant electron dispersion~(\ref{eq:6}) in the foam due to charge conservation was considered. The anomalous terms expected to appear in calculation and being proportional to the ratio $k/\Md$, are highly suppressed~(and hence significantly smaller than~(\ref{eq:21})) in this case, as the momentum $k\simeq 0$ which implies $\delta E_{D}\simeq 0$. The effective formation length then coincides with the standard one $\lambda_{f}\simeq\frac{2E(E-\omega)}{m_{e}^{2}\omega}$, while this result can be modified by multiple scattering and other suppression factors such that the infrared divergence is removed.

While for photons at moderate energies, the process may be mildly affected, nevertheless it is important to note that in order for an electron to emit a real photon as it interacts with the field of the nucleus, part of its surrounding virtual photon field shakes loose, a case for which no modification is induced in this scenario. As we have explained, LV effect within such D-brane models is due to multiple interactions between photons and D-particles, hence cannot be relevant in any mere process of photon production. This is another difference from the phenomenological consideration of~\cite{Vankov:2002gt}, where due to electron LV effect, bremsstrahlung is strongly suppressed, similarly to pair creation on nuclei by photons, reducing the formation length $\lambda_{f}$ from~(\ref{eq:21}), and hence the probability for radiation. There is currently no direct limit from bremsstrahlung though as the latter is usually viewed as unmodified, for practical~(LV constraining) purposes, as also mentioned before.

\section{Instead of conclusions}

Our modest proposal~\cite{Li:2021gah,Li:2021tcw,Li:2022sgs} that a possible signature of light-speed variation in cosmic photons could be attributed to propagation in a stochastic QG D-foam medium, stimulated us to examine effects induced in atmospheric showers of these photons. The existence of ``shower anomalies'' has been previously suggested based on general form of Planck scale-motivated LV dispersion relations which characterize the effective field theory model of LV in~\cite{Myers:2003fd,Satunin:2023yvj}. Using the concept of radiation formation length we have shown, however, that such anomalies may be eliminated in the present model. Qualitatively, this difference has been traced to the fact that the assumption of energy conservation implicit in previous analyses of scattering kinematics does not hold in foamy situations. Our lack of knowledge of D-foam/matter interaction dynamics for which presently one does not have analytic control -- a major limitation for conclusions drawn from above -- is of much smaller impact as we did tentative calculations while such aspects nevertheless deserve future investigation.

In any rate, as a cautionary theoretical note, one always needs a concrete formalism to study possible effects for the relativistic kinematics. This is in contrast to the result~\cite{Vankov:2002gt} that both the deformed velocity of light and the suppressed photon showers can all be induced by the same subluminal photon dispersion~(\ref{eq:1}). This underlines the limitations of a purely phenomenological model: without explicit guidance from a QG theory it is difficult to be certain of the possible LV in different observational contexts.

Note that while another process, the direct photonuclear reaction, can also be $\g$-rays' first reaction in air, it is much inefficient than BH ones even for photon at $10^{19}$~eV, which has not been detected yet. Still, we want to give some clues to the problem of whether it and neutrino--nucleus interaction are affected, since at least the latter is relevant for detection of LV cosmic neutrinos~(\ref{eq:3})~\cite{Huang:2018ham,Huang:2019etr,Huang:2022xto}, at IceCube~\cite{IceCube:2013cdg}, a Cherenkov detector. We note, again, that it is a prerequisite of foam corrections that there is nontrivial interaction between the incident particles with D-foam. The fact that atomic nucleus consists of protons and neutrons which are not structureless particles complicates matters. Although the neutrons are supposed to interact with this foam, their quark substructure may suppress this by introducing additional~(and significant) suppression factors in the pertinent QG parameter, due to electric charge of the latter, thereby leading to hardly modified neutrino~(or photon) scattering off a nucleus.

We should emphasize, however, that at the present level of understanding of the model(s) describing the recoil foam backgrounds one is unable to rigorously establish that neutrino initiated cascade within the detector material is unaffected~(or, negligibly affected) by the new features. But, as discussed, the nonelementary quark and gluon structure of nuclei does play an important suppressing role of any foam effects. Since the D-foam scenario addressed here has been found to introduce rather smooth departures from relativity~(smoother than in a LV theory), it is natural to expect, that nothing much would change for such kind of phenomena involving particles of a few $10^{15}$~eV and even the recent 220~PeV KM3NeT neutrino~\cite{KM3NeT:2025npi}~(still small in comparison to the scale characteristic of any QG-motivated effects for a detector). In fact Lorentz symmetry is restored \emph{on average} in this type of models by the zero mean of the distribution of $r$ despite the presence of recoil $u\sim\gs r E/\Ms$, as pointed out elsewhere~\cite{Li:2022sgs} for other purposes.

Obviously, at least, the energy loss of charged leptons via Cherenkov effects inside a detector remains intact, since as explained, foam interactions with such electrically charged elementary constituents of matter are absent on account of charge conservation. The above discussion seems to justify at this stage a general theorem, stating that only reactions with particles susceptible to the D-brane foam in the \emph{initial} state~(i.e., particle reactions exclusively involving photons, or neutrinos, for which there is no obstacle to interact with D-particles) are predominantly affected in the model. This provides thus additional reasons for suppressed corrections to bremsstrahlung from the foam contributions.

A final remark is that nonconservation of particle energy that is due to background changes during particle interactions in D-brane foam does not make its appearance during propagation, where momentum of an observable particle is exactly conserved~\cite{Kogan:1996zv,Mavromatos:1998nz}, following the closure of a worldsheet logarithmic algebra, and, its energy conserved on the average~\cite{Ellis:2000sf}~(see also first review in~\cite{Li:2025stf}). For photon \emph{propagation} in the vacuum of D-foam space-time, one would not expect any uncertainties in energies and momenta to dominate over the modifications of the dispersion relations~(\ref{eq:4}). This is to be contrasted with foam cases like that of~\cite{Amelino-Camelia:2002qcs,Basu:2005kt} where energy--momentum fluctuations dominate the propagation and are bounded by analyzing cosmic-ray data. In our approach, energy violation is merely relevant for \emph{interactions} of $\g$-rays with atmospheric nuclei as discussed here or with cosmic background fields. The foam impact on the latter reaction is up for debate, and, as already mentioned, is left for future study.

The main conclusion from this Letter, therefore, is that, in the string model of quantum foam considered here, subluminal modified photon dispersion relations~(\ref{eq:4}) that may be compatible with a phenomenological hint of light-speed variation at a few $10^{17}$~GeV~\cite{Shao:2009bv,Zhang:2014wpb,Amelino-Camelia:2016ohi,Xu:2016zxi,Xu:2016zsa,Xu:2018ien,Liu:2018qrg,Chen:2019avc,Zhu:2021pml,Zhu:2022usw,Song:2024and,Song:2025qej,Song:2025myx,Song:2025akr} can evade drastic modifications to pair creation and bremsstrahlung under certain conditions. Thus, LV with such intensities may concordant in the context of such models with tests of subluminal photons in giant air showers. Otherwise the strong suppression effects induced in other LV proposals that are not observed experimentally in shower formation would allow~\cite{Satunin:2023yvj} to imply trans-Planckian limits for the relevant scale. A detailed estimate of more complicated effects of foam from dynamical aspect is anticipated to refine the above analysis, which might help better probing or constraining this type of QG models from string theory.

\section*{Acknowledgments}

This work was supported by the China Postdoctoral Science Foundation~(CPSF) under Grant No. 2024M750046. The financial support received from the Postdoctoral Fellowship Program~(Grade B) of CPSF under the Grant No. GZB20230032 and from a \emph{Boya} Postdoctoral Fellowship of Peking University is acknowledged. The author would like also to thank the COST Action CA23130 --- Bridging high and low energies in search of quantum gravity~(BridgeQG).

\section*{Note added}

A potential new line of evidence for the light-speed variation~\cite{Shao:2009bv,Zhang:2014wpb,Amelino-Camelia:2016ohi,Xu:2016zxi,Xu:2016zsa,Xu:2018ien,Liu:2018qrg,Chen:2019avc,Zhu:2021pml,Zhu:2022usw,Song:2024and,Song:2025qej,Song:2025myx,Song:2025akr} was recently reported~\cite{Song:2025ksi}, based on the detection of a 300~TeV photon-like air shower event by Carpet--3 coincident with GRB~221009A, offering further support for the theory here championed.


\end{document}